\documentstyle[prd,preprint,aps]{revtex}

\begin{document}

\reversemarginpar
\tighten

\title{Kerr black hole as a quantum rotator}

\author{Gilad Gour\thanks{E-mail:~gilgour@phys.ualberta.ca}
and A.J.M. Medved\thanks{E-mail:~amedved@phys.ualberta.ca}}

\address{
Department of Physics and Theoretical Physics Institute\\
University of Alberta\\
Edmonton, Canada T6G-2J1\\}

\maketitle

\begin{abstract}
It has been proposed by Bekenstein and others that the horizon
area of a black hole conforms, upon quantization,
to a discrete and uniformly spaced spectrum. 
In this paper, we consider the area spectrum
for the highly non-trivial case of a rotating
(Kerr) black hole solution. Following a prior
work by Barvinsky, Das and Kunstatter, we are
able to express the area spectrum in terms of
an integer-valued quantum number and  an
angular-momentum operator. Moreover, by using an analogy
between the Kerr black hole and a quantum rotator,
we are able to quantize the angular-momentum sector.
We find the area spectrum to be 
$A_{n,J_{cl}}=8\pi\hbar(n+J_{cl}+1/2)$,
where $n$  and $J_{cl}$ are both integers.
The quantum number $J_{cl}$ is related to
but distinct from the eigenvalue $j$ of the angular momentum
of the black hole. Actually, it represents the ``classical'' angular 
momentum and, for $J_{cl}\gg 1$, $J_{cl}\approx j$.

\end{abstract}

\section{Introduction}

As is well known since the early seventies, 
black holes  behave dynamically as thermodynamic
systems~\cite{BEK,HAW}. In particular, 
 the surface area ($A$) of the horizon   plays the role
of the entropy ($S$) and the surface gravity ($\kappa$) at the horizon
serves as the temperature ($T$); that is: 
\begin{equation}
S= {A \over 4 l_p^2}\quad\quad and \quad\quad T={\kappa\over 2\pi}.
\label{0.1}
\end{equation}
(Here and throughout, the spacetime dimensionality is four,
 $l_p^2\sim\hbar$ is the Planck constant and the fundamental constants
 $c$, $G$, $k_B$ have been set equal to unity.)
Thanks to Hawking's discovery that quantum black holes radiate
at precisely the above value of temperature~\cite{HAW2},
this thermodynamic analogy has since been elevated to the status of a 
physical theory.

One of the outstanding open questions in gravitational theory
is the microscopic origin of this thermodynamic behavior.
In all likelihood, such a question can  only be
truly resolved in the context of a quantum theory of    gravity;
 a theory for which our understanding is conspicuously
incomplete.  Nonetheless, there
are still  fundamental issues that can  be addressed
even in the absence of the full-fledged quantum theory.
One such  question  is what is the quantum spectrum  
of the black hole observables?

That the black hole horizon area, in particular, should
be quantized was first argued for by
Bekenstein~\cite{BEK2} (also see~\cite{MUK,BEKX}).
The support for this argument comes from
 the observation that $A$
behaves, for a slowly changing black hole, as an adiabatic invariant 
\cite{BEK3}. It is significant that, as Bekenstein pointed out,
a classical adiabatic invariant corresponds to a quantum
observable with a discrete spectrum, by virtue of Ehrenfest's
principle.

On quite general grounds, Bekenstein has  suggested
the following explicit form for the area spectrum
~\cite{BEK2,MUK}:
\begin{equation}
A=\epsilon l_p^2 n, \quad\quad\quad n=0,1,2,...,
\label{0.2}
\end{equation}
where $\epsilon$ is a numerical factor of the order unity.
(Note that a non-vanishing but positive zero-point term may also
be considered.)
The crucial point in this formulation is the equal spacing between
the levels. This  can be viewed as a consequence of
the uncertainty principle, as a quantum point particle
cannot be localized better than one Compton length,
and this naturally leads to a minimal increase in the
horizon area of $(\Delta A)_{min}=\epsilon l_p^2$~\cite{BEK2,BEKX}.

Since the original heuristic arguments of Bekenstein, there has been
a substantial amount of work in trying to derive 
the spectrum (\ref{0.2}) by more rigorous means (see
\cite{BEKG} for a  list of relevant references). An example of
a more rigorous proof of the equally spaced area spectrum, as well as the
degeneracy of the area levels, can be found in the algebraic approach to black 
hole quantization~\cite{Ann_Arbor,BEKG,GOUR}.

Of particular relevance to the upcoming analysis is
a program that was initiated by Barvinsky and Kunstatter~\cite{BG}.
Their methodology is based on expressing the black
hole dynamics in terms of a  reduced phase space\footnote{To achieve
the desired form of phase space, one requires  a midisuperspace
type of  approximation - for instance, by imposing spherical symmetry - 
so as to sufficiently  reduce the number of 
black hole degrees of freedom.}
 and then
applying an appropriate process of quantization. 
For a static,  uncharged black hole,
this phase space consists of only the black hole mass 
observable and its canonical conjugate~\cite{KUN,KUC}.
(This simplicity can be viewed as a manifestation
of either Birkhoff's theorem~\cite{WAL} or
the ``no-hair'' principles of black holes~\cite{NH}.)
One vital assumption was required in this analysis;
 namely, the authors assumed that the
conjugate to the mass is periodic over an interval
of $2\pi/\kappa$. They did, however, justify this input
by way of Euclidean considerations. (We elaborate
on the logistics of this point later on in the paper.)
Ultimately, the area spectrum (\ref{0.2}) was indeed reproduced
with the particular value of $\epsilon=8\pi$ 
(and a zero-point contribution of $4\pi l_p^2$).

The general procedure of~\cite{BG} was later extended
by Barvinsky, Das and Kunstatter
to the case of a charged but still static black hole~\cite{BDG}
(also see~\cite{BDGX}). In this case, the reduced phase
space now consists of the two relevant observables 
(the mass and the charge, $Q$) and their respective conjugates
\cite{LMG}. Assuming the same periodicity condition as before,
the authors found  the following for the area spectrum:
\begin{equation}
A-A_{ext}(Q)= 8\pi l_p^2 \left(n+  {1\over 2}\right),
\label{0.3}
\end{equation}
where  $A_{ext}(Q)$ is the extremal value of the horizon 
area\footnote{Note that a
charged or rotating black hole typically has a pair of
distinct horizons,  with their coincidence  determining
the point of extremality. Further note that, throughout
this paper, an unqualified $A$ always signifies the
area of the outermost horizon.}
(expressed as a function of the charge).
Significantly, this extremal value represents, for a given value
of $Q$, a lower bound on the horizon area of
a classical black hole. Note, however,
that because of the zero-point term in Eq.(\ref{0.3}),
the quantum black hole can not approach this extremal
value. (The authors of~\cite{BDG} attributed this
censoring feature to  the effects of quantum fluctuations.)

Barvinsky {\it et al} went on to quantize the
charge sector of the theory and ultimately found that~\cite{BDG}
\begin{equation}
A=8\pi l_p^2 \left(n+ {p\over 2} +{1\over 2}\right), 
\quad\quad\quad n,p=0,1,2,...,
\label{0.4}
\end{equation}
where the ``new'' quantum number $p$ is related to the
black hole charge  according to $Q^2=\hbar p$.

The objective of the current paper is to further extend
the above program to the case of a rotating black hole.
(We will be assuming, for sake of simplicity, 
an uncharged black hole and always a four-dimensional 
spacetime.)
This seems an {\it a priori} difficult task, given
that there is no rigorous evidence that a rotating
black hole can be described by an analogously simple
form of reduced phase space.
Nonetheless, we argue that, on the basis of
the ``no-hair'' principles~\cite{NH}, that this
should indeed be the case, with the relevant
observables in the phase space now being the mass and
an angular-momentum vector.  The latter inclusion necessitates
six additional degrees of freedom; for instance,
the three Cartesian components of the angular momentum and their respective
conjugates. (However, it will be shown later that 
the choice of Cartesian components
is inappropriate and we will work, instead, with 
the Euler components as the initial basis.) Let us emphasize
that this conjectural form of reduced phase space and
the periodicity constraint on the conjugate to
the mass~\cite{BG} are the {\it only} assumptions used in the following
analysis. (Also note that, later on, we will provide additional, independent
support for this periodicity constraint.) 

Before discussing the contents of this paper, let us point out that
the area spectrum of a rotating black hole has
recently been considered by Makela {\it et al}~\cite{MRLP}
(also see~\cite{LM} for earlier studies
on  static black holes). 
Their approach, which differs substantially from
that of Barvinsky {\it et al}, is based on formulating
a Schrodinger-like equation for the black hole
observables and  quantizing this equation via
a WKB  analysis. Even without bringing rotation
into the discussion, the results of~\cite{MRLP} are somewhat different
than those discussed above.
For instance, the spacing between levels was found
to be  $\epsilon=32\pi$ (translated to our
notation), and the quantity being quantized is not
$A-A_{ext}$  but rather $A + A_{-}$ (where
$A_{-}$ represents the area of the inner black hole horizon).
This latter
distinction makes a direct comparison between
the two approaches rather non-trivial. 

The remainder of the paper is organized as follows. In 
the next section, we consider some relevant properties,
at the classical level,
of a rotating (Kerr) black hole. We then propose
a reduced phase space and transform it into
a form that is suitable for the subsequent quantum analysis.
In Section 3,  following the
general methodology  of  Barvinsky {\it et al}~\cite{BDG},
we are able to quantize the reduced phase space.
This eventually  yields  an expression for the area spectrum
in a form which is analogous to that of  Eq.(\ref{0.3}).
In Section 4, we focus on the angular-momentum sector,
and demonstrate that  the  spin eigenvalues are
necessarily restricted to taking on integer 
values. In this way, we are able to derive an
explicit, unambiguous form of the area spectrum,
which is clearly evenly spaced and behaves as
intuitively expected in the limiting cases of interest.
The final section contains a summary.

\section{Classical Analysis}

Let us begin here by considering the physically relevant model
of interest; namely, a four-dimensional spacetime
containing a rotating black hole.  In this analysis,  we will
focus on the  Kerr black hole, which may be regarded as
the most general solution of the vacuum Einstein equations
with vanishing electrostatic charge.
In this particular section, considerations 
will be restricted to the classical level.

Thanks to the ``no-hair'' principles of black holes~\cite{NH},
we are safe in assuming that an external observer
can describe
 the system 
strictly in terms of a few macroscopic parameters;
in particular, the black hole mass, $M$, and an angular
momentum, $\vec{J}_{cl}$.\footnote{We include a subscript 
on this {\it classical} form of the angular momentum so as
to avoid confusion in the later analysis.} Moreover,
the well-known first law of black hole mechanics~\cite{BEK,HAW}
relates these quantities in the following manner:
\begin{equation}
dM={\kappa\over 8\pi}dA+\Omega dJ_{cl}.
\label{1}
\end{equation}
Here, 
$A$ is the (outermost) horizon area, $\kappa$ is
the surface gravity at this horizon,
$\Omega$ is the angular velocity
of the black hole, 
and $J_{cl}=|\vec{J}_{cl}|$ is the magnitude
of the angular-momentum vector.

For the black hole of interest, the above quantities
are explicitly known~\cite{WAL}:
\begin{equation}
A=8\pi M\left[M+\sqrt{M^2-{J_{cl}^2\over M^2}}\right]
\label{2}
\end{equation}
or equivalently:
\begin{equation}
M^2={A\over 16\pi}+4\pi{J_{cl}^2\over A},
\label{3}
\end{equation}
and:
\begin{equation}
\kappa=\left.8\pi{\partial M\over \partial A}\right|_{J_{cl}}=
{1\over 4M}-16\pi^2{J_{cl}^2\over MA^2},
\label{4}
\end{equation}
\begin{equation}
\Omega=\left.{\partial M\over \partial J}\right|_{A}
=4\pi{J_{cl}\over AM}.
\label{5}
\end{equation}

Extrapolating the well-understood dynamics of static black holes
\cite{KUN,KUC,LMG},
we will assume  that any
classical  black hole 
can be described (by an external observer) in terms of a reduced phase
space consisting of the physical observables
and their respective canonical conjugates.
(For a relevant discussion in the context of
rotating black holes, see~\cite{MRLP}.)
Focusing on the current scenario,  one
might be inclined to describe the  reduced phase
space  in terms of $M$, $J_{x}$, $J_{y}$ and
$J_{z}$ (where $J_{x}$, {\it etc.} are the usual
angular-momentum components in Cartesian coordinates).
However, these variables are actually a poor choice
because of their failure to commute 
(in terms of Poisson brackets). 
Therefore, the set $M$, $J_{x}$, $J_{y}$ and
$J_{z}$ cannot be considered as a set of {\it generalized 
coordinates}.
We can, however,
rectify this situation by alternatively considering 
the {\it Euler} components~\cite{EUL} of the angular 
momentum: 
\begin{equation}
J_{\alpha} , J_{\beta}, J_{\gamma},
\label{6}
\end{equation}
along with their respective conjugates, the three 
{\it Euler} angles,
$\alpha$, $\beta$ and $\gamma$. The Cartesian components of 
the angular momentum can be written in terms of the Euler
components~\cite{EUL}:
\begin{eqnarray}
J_{x} & = & -\cos\alpha\cot\beta J_{\alpha}
-\sin\alpha J_{\beta} +\frac{\cos\alpha}{\sin\beta}J_{\gamma}\nonumber\\
J_{y} & = & -\sin\alpha\cot\beta J_{\alpha}
+\cos\alpha J_{\beta} +\frac{\sin\alpha}{\sin\beta}J_{\gamma}\nonumber\\
J_{z} & = & J_{\alpha}   
\label{rrr}
\end{eqnarray}

If we adopt the common-sense assumption that
the horizon area is invariant under rotation,
it is  clear that
\begin{eqnarray} 
&\;& A,\;J_{\alpha},\;J_{\beta},\;J_{\gamma}, 
\label{set1}\\ 
&\;& P_{A},\;\alpha,\;\beta,\;\gamma, 
\label{set11}
\end{eqnarray}
forms the desired set of generalized (commuting) 
coordinates~(\ref{set1}) 
and their canonical conjugates~(\ref{set11}). 
However, we would like to work with
a set that includes $M$ because, later on, 
the periodicity of its conjugate, $P_{M}$, will  be exploited
in order to obtain the area spectrum.

The set
$$
M,\;J_{\alpha},\;J_{\beta},\;J_{\gamma}, 
$$
on the other hand, is a poor choice because
\begin{equation}
\{M,J_{\beta}\}\neq 0,
\label{noteq}
\end{equation}   
where $\lbrace\quad,\quad \rbrace$  denotes a commutator
(Poisson) bracket\footnote{The derivatives
are taken with respect to the generalized coordinates  
in~(\ref{set1}) and their canonical conjugates~(\ref{set11}).} 
in the Dirac sense~\cite{DIR}. To prove Eq.(\ref{noteq}),
it is enough to show that $J_{cl}$ does not commute with
$J_{\beta}$ ({\it cf}, Eq.(\ref{3})). This can, in fact, be seen from the 
explicit expression for $J_{cl}$:
\begin{eqnarray}
J_{cl}^2&=&J_{x}^2+J_{y}^2+J_{z}^2
\nonumber \\
&=& {1\over \sin^2\beta}\left[J_{\alpha}^2+J_{\gamma}^2-2\cos\beta
J_{\alpha}J_{\gamma}\right]+J_{\beta}^2,
\label{14}
\end{eqnarray} 
where, in this section, we treat $J_{\alpha}$, $J_{\beta}$, $J_{\gamma}$ 
as classical ({\it i.e.}, non-operating) quantities.
Note the presence of $\beta$ in the above relation,
as this clearly demonstrates that $\{J_{cl},J_{\beta}\}\neq 0$.

Eq.(\ref{14}) also shows  that $J_{cl}$ commutes with both $J_{\alpha}$
and $J_{\gamma}$.  This prompts us to introduce a new set of variables: 
\begin{equation}
M=M(A,J_{cl}), J_{cl} , J_{\alpha} , J_{\gamma} ,
\label{15}
\end{equation}
along with their {\it hypothetical} conjugates:
\begin{equation}
\Pi _{M}, \Pi _{cl}, \Pi _{\alpha}, \Pi _{\gamma}.
\label{16}
\end{equation} 
At this point, we use the qualifier  ``hypothetical'',
as it is not {\it a priori} clear that there exists
a transformation from Eqs.(\ref{set1},\ref{set11}) to
Eqs.(\ref{15},\ref{16}) that is truly canonical.
To be explicit, such a transformation requires that
\begin{equation}
\lbrace M,P_{M} \rbrace = \lbrace J_{cl},P_{cl} \rbrace
=\lbrace J_{\alpha},P_{\alpha} \rbrace 
= \lbrace J_{\gamma},P_{\gamma} \rbrace
=1, 
\label{17}
\end{equation}
\begin{equation}
\lbrace all\quad other\quad combinations \rbrace=0,
\label{18}
\end{equation}
where for arbitrary $\mu$ and $\nu$:
\begin{eqnarray}
\lbrace{\mu,\nu}\rbrace ={\partial \mu\over \partial A}{\partial \nu\over\partial P_{A}} 
- {\partial \mu\over \partial P_{A}}{\partial \nu\over\partial A} 
+ {\partial \mu\over \partial J_{\alpha}}{\partial \nu\over\partial \alpha}  
- {\partial \mu\over \partial \alpha}{\partial \nu\over\partial J_{\alpha}}
\nonumber\\ 
+ {\partial \mu\over \partial J_{\beta}}{\partial \nu\over\partial \beta} 
- {\partial \mu\over \partial \beta}{\partial \nu\over\partial J_{\beta}} 
+ {\partial \mu\over \partial J_{\gamma}}{\partial \nu\over\partial \gamma} 
- {\partial \mu\over \partial \gamma}{\partial \nu\over\partial J_{\gamma}}.
\label{19}
\end{eqnarray}

As it so happens, the canonical transformation
in question does indeed exist, 
as can be shown in two steps. First,
we make a {\it canonical} transformation from 
Eqs.(\ref{set1},\ref{set11}) to the set:
\begin{eqnarray} 
&\;& A,\;J_{cl},\;J_{\alpha},\;J_{\gamma}, 
\label{set01}\\ 
&\;& P_{A},\;P_{cl},\;P_{\alpha},\;P{\gamma}, 
\label{set011}
\end{eqnarray} 
where we have exchanged $J_{\beta}$ with $J_{cl}$.
Then, after some
lengthy but straightforward calculations, one can verify
that Eqs.(\ref{17},\ref{18}) are consistently satisfied 
with the following conjugates:
\begin{eqnarray}
\Pi _{M} & = & {8\pi\over \kappa} P_{A},
\label{20}\\
\Pi _{cl} & = & -{8\pi\over\kappa}\Omega P_{A} +P_{cl},
\label{21}\\
\Pi _{\alpha} & = & P_{\alpha},
\label{22}\\
\Pi _{\gamma} & = & P_{\gamma}.
\label{23}
\end{eqnarray}

\section{Quantizing the Area}

With the black hole mass ($M$) and its conjugate
($\Pi _{M}$) contained within the reduced phase space,
we are now well positioned to begin a process
of quantization  in the manner of Barvinsky {\it et al}~\cite{BDG}.
In following the prescribed methodology,
we must necessarily   invoke the following condition of periodicity:
\begin{equation}
\Pi _{M}=\Pi _{M}+{2\pi\over\kappa}.
\label{24}
\end{equation}
Although an assumption, this condition follows quite
naturally from a pair of observations.
{\it (i)} The conjugate to the mass, $\Pi _{M}$, can be identified 
with the time separation at infinity~\cite{KUC}; that is,
$\Pi _{M}$ directly measures the difference in Schwarzschild-like
time between the ends of a spacelike slice that extends across
the relevant Kruskal diagram. {\it (ii)} In the Euclidean
(or imaginary time) sector of a black hole spacetime,
the Schwarzschild-like time is periodic~\cite{GH}, with the period
given precisely by $2\pi/\kappa$.
 
At least naively, these two observations, when take together, 
suggest that  $\Pi _{M}$ should be constrained with  the 
specified periodicity.  
On the other hand,
the first observation follows from a purely  Lorentzian
perspective (Kruskal coordinates extend over the
entire Lorentzian spacetime, whereas Euclidean coordinates
reduce the black hole interior to a single point),
and so it is unclear if  {\it i} can be translated into
the Euclidean framework  of {\it ii}.  
For this reason,  the above condition should, at this point, 
be regarded as a well-motivated but conjectural input. 
For further justification and related discussion,
 see~\cite{BDG} (especially, pages 15-16 in the archival version).
We also provide, in the next section, an independent argument 
that further substantiates the validity of Eq.(\ref{24}). 

Again following~\cite{BDG}, let us now introduce
a new pair of variables that directly incorporate the 
periodic nature of $\Pi _{M}$:
\begin{equation}
X=\sqrt{B(M,J_{cl},J_{\alpha},J_{\gamma})\over\pi}\cos(\Pi _{M}\kappa),
\label{25}
\end{equation}
\begin{equation}
{\cal P}_{X}=\sqrt{B(M,J_{cl},J_{\alpha},J_{\gamma})\over\pi}\sin(\Pi _{M}\kappa).
\label{26}
\end{equation}
Here, we have included a yet-to-be-determined
function, $B$, of the phase-space observables.\footnote{Note that,
as written above, 
$B$ has units of area; that is, $B\sim\hbar$.}
The underlying  premise is that $B$ can
be (at least partially) fixed with the  constraint that 
Eqs.(\ref{15},\ref{16}) transform {\it canonically} into
the set of observables:
\begin{equation}
X,J_{cl},J_{\alpha},J_{\gamma}
\label{27}
\end{equation}
and their conjugates: 
\begin{equation}
{\cal P}_{X},{\cal P}_{cl},{\cal P}_{\alpha},{\cal P}_{\gamma}.
\label{28}
\end{equation}

With the above in mind, let us consider the
following necessary and sufficient condition for
a canonical transformation:
\begin{equation}
{\cal P}_{X}\delta X + {\cal P}_{cl}\delta J_{cl} + {\cal P}_{\alpha}
\delta J_{\alpha} +{\cal P}_{\gamma}\delta J_{\gamma}
= \Pi _{M}\delta M + \Pi _{cl}\delta J_{cl} +\Pi _{\alpha}\delta J_{\alpha} 
+\Pi _{\gamma}\delta J_{\gamma}.
\label{29}
\end{equation}
Up to a total variation, it can be shown that
\begin{equation}
{\cal P}_{X}\delta X={\kappa\Pi _{M}\over 2\pi}\left[
{\partial B\over \partial M}\delta M + 
{\partial B\over \partial J_{cl}}\delta J_{cl} + 
{\partial B\over \partial J_{\alpha}}\delta J_{\alpha} + 
{\partial B\over \partial J_{\gamma}}\delta J_{\gamma} \right].
\label{30}
\end{equation} 
Substituting Eq.(\ref{30}) into Eq.(\ref{29}), we
are then able to deduce the following:
\begin{equation}
{\partial B\over\partial M}= {2\pi\over\kappa},
\label{31}
\end{equation}
\begin{equation}
{\partial B\over\partial J_{cl}}= {2\pi\over\kappa\Pi _{M}}
\left(\Pi _{cl}-{\cal P}_{cl}\right),
\label{32}
\end{equation}
\begin{equation}
{\partial B\over\partial J_{\alpha}}= {2\pi\over\kappa\Pi _{M}}
\left(\Pi _{\alpha}-{\cal P}_{\alpha}\right),
\label{33}
\end{equation}
\begin{equation}
{\partial B\over\partial J_{\gamma}}= {2\pi\over\kappa\Pi _{M}}
\left(\Pi _{\gamma}-{\cal P}_{\gamma}\right).
\label{34}
\end{equation}

It is informative to compare Eq.(\ref{31}) with Eq.(\ref{4}),
which immediately indicates that $\partial A/\partial M=4 \partial B/\partial M$.
Hence, we can write
\begin{equation}
B(M,J_{cl},J_{\alpha},J_{\gamma})={1\over 4}A(M,J_{cl})+F(J_{cl},J_{\alpha},J_{\gamma}),
\label{35}
\end{equation}
where $F$ is an essentially arbitrary function
of the angular momentum. That is to say, for any
well-behaved choice of $F$, one will always be
able to find expressions for ${\cal P}_{cl}$, ${\cal P}_{\alpha}$ and 
${\cal P}_{\gamma}$ that satisfy Eqs.(\ref{32}-\ref{34}).

In spite of this freedom in choosing $F$, there
is only one particular form that will be useful
for the quantization of the area\cite{BDG}.
First, it is relevant that, regardless of
the choice of $F$, the function $B$ is bounded
from below. This follows from the lower bound
that exists on the area, $A$. To be precise,
for a rotating  black hole,
$A$ can not, classically, fall below its
{\it extremal} value.\footnote{This realization follows
from the censorship of  naked singularities,
which is usually assumed to be the case~\cite{WAL}.}
 This occurs when $M^2=J_{cl}$ 
({\it cf}, Eq.(\ref{2})), and so:
\begin{equation}
A \geq A_{ext}= 8\pi J_{cl}.
\label{36}
\end{equation}
As elaborated on below, it turns out to be convenient
if $F$ is chosen so that Eq.(\ref{36}) translates
into $B\geq 0$.  Following this prescription,
we can unambiguously  set $F=-8\pi J_{cl}/4$  and thus obtain
\begin{equation}
B={1\over 4}\left[A(M,J_{cl})-8\pi J_{cl}\right].
\label{37}
\end{equation}

Let us now recall Eqs.(\ref{25},\ref{26}), which
can be squared and summed to yield $B=\pi(X^2+{\cal P}_{X}^2)$.
Hence, Eq.(\ref{37}) can be suggestively re-expressed as follows:
\begin{equation}
X^2+{\cal P}_{X}^2={1\over 4\pi}\left[A(M,J_{cl})-8\pi J_{cl}\right]\geq 0.
\label{38}
\end{equation}
In this way, we have mapped  the mass and its conjugate,
$M$ and $\Pi _{M}$, into a {\it complete} two-dimensional plane,
$X$ and ${\cal P}_{X}$. Any other choice of $F$ would have left
a ``hole'' in this plane and complicated 
the prospective quantization  with the need for non-trivial boundary 
conditions.

Next, let us elevate any classically defined quantity in Eq.(\ref{38})
to the status of a quantum operator. Adopting the conventional
``hat'' notation, we then have
\begin{equation}
\frac{1}{2\pi}\hat{B}\equiv
{1\over 8\pi}\left[\hat{A}-8\pi\hat{J}_{cl}\right]
={\hat{X}^2\over 2}+{\hat{\cal P}_{X}^2\over 2}.
\label{39}
\end{equation}
Since the domain of $\hat{X}$ and $\hat{\cal P}_{X}$ is  
an entire two-dimensional
plane, the quantization of the right-hand side becomes trivial. 
Indeed, the spectrum is readily identifiable with
 that of a harmonic oscillator, and so:
\begin{equation}
\frac{1}{2\pi}B_{n}=8\pi\hbar\left(n+{1\over 2}\right),
\quad\quad\quad n=0,1,2,...,
\label{40}
\end{equation}
where $B_{n}$ are the eigenstates of the operator $\hat{B}$.

Our task is not, of course, complete until the spectra for
$\hat{A}$ and $\hat{J}_{cl}$ have been explicitly separated.
There is, however, an interesting observation
that can be made without any further analysis. 
Namely, we can see from Eq.(\ref{40}) that quantum fluctuations
will always prevent the rotating black hole
from ever reaching a precise state of extremality
(since the right-hand side can never quite vanish).
This result can best  be viewed as 
a quantum black hole version of the
 third law of  thermodynamics.
Note that
a similar observation was also made for charged (non-rotating)
black holes in the prior work of Barvinsky {\it et al}~\cite{BDG}.

\section{Quantizing the Angular Momentum}

Since our principle objective is to find the area spectrum for
a rotating  black hole,  the  preceding outcome (\ref{40})
emphasizes the importance in knowing the spectrum of $\hat{J}_{cl}$. 
Fortunately, it turns out that the spectrum of $\hat{J}_{cl}$ can be 
obtained by way of some simple calculations.

To proceed in the stated direction, let us first take
note of the operator form of this
angular momentum ({\it cf}, Eq.(\ref{14})):
\begin{equation}
\hat{J}_{cl}^2 
= {1\over \sin^2\beta}\left[\hat{J}_{\alpha}^2+\hat{J}_{\gamma}^2-2\cos\beta
\hat{J}_{\alpha} \hat{J}_{\gamma}\right]+\hat{J}_{\beta}^2.
\label{41}
\end{equation}
In the above, the order of the operators in each element   
is not important because $\beta$, $\hat{J}_{\alpha}$ 
and $\hat{J}_{\gamma}$  all commute with each other. Therefore, the 
transition from the classical $J_{cl}$ to the quantum $\hat{J}_{cl}$ is 
well defined.
Furthermore, it should be kept in mind that this expression
is obtained by {\it first} summing the squares of the
individual components ($J_{x}$, $J_{y}$, $J_{z}$) and
{\it then} quantizing. This makes $\hat{J}_{cl}$ distinct  
from the ``traditional'' quantum operator, $\hat{J}$,
which is obtained by first quantizing the components
and then summing the squares. The importance of this
distinction will become evident below. 

In order to work with the quantum Euler components of the angular
momentum, we will employ the usual identification of the operators, 
$\hat{J}_{\eta}=-i\hbar\partial/\partial\eta$ 
(for any component $\eta=\alpha,\;\beta$
or $\gamma$). Usefully, it can be shown~\cite{EUL} that Eq.(\ref{rrr}) 
remains valid
when the classical components of the angular momentum are replaced by
these quantum operators. 
Therefore, the square of the conventional angular momentum
is given by
\begin{eqnarray}
\hat{J}^2&=&\hat{J}_{x}^2+\hat{J}_{y}^2+\hat{J}_{z}^2
\nonumber \\
&=& {1\over \sin^2\beta}\left[\hat{J}_{\alpha}^2
+\hat{J}_{\gamma}^2-2\cos\beta
\hat{J}_{\alpha}\hat{J}_{\gamma}\right]+\hat{J}_{\beta}^2
-i\hbar\cot\beta\hat{J}_{\beta}.
\label{48}
\end{eqnarray}
That is ({\it cf}, Eq.(\ref{41})):
\begin{equation}
\hat{J}_{cl}^2-\hat{J}^2=\hbar^2\cot\beta{\partial\over \partial\beta}.
\label{49}
\end{equation}
Hence, the spectrum of $\hat{J}_{cl}^2$ must be different than 
$\hbar^2 j(j+1)$.
 
Since both $\hat{J}$ and $\hat{J}_{cl}$ commute with both of 
$\hat{J}_{\alpha}$ 
and $\hat{J}_{\gamma}$, there are two natural sets of
angular-momentum eigenstates:
the conventional set  
$\{|j,m_{\alpha},m_{\gamma}\rangle\}$ (where $j=0,1/2,1,...$
and $m_{\alpha},m_{\gamma}=-j,-j+1,...,j$)\footnote{Note
that the degeneracy of the angular momentum is $(2j+1)^{2}$,
just as it appears in quantum rotators~\cite{EUL}.}
and 
$\{|J_{cl},m_{\alpha},m_{\gamma}\rangle\}$ (with $\{J_{cl}\}$ being
the eigenvalues of $\hat{J}_{cl}$). The first basis
is able to diagonalize
simultaneously $\hat{J}$, $\hat{J}_{\alpha}$ and $\hat{J}_{\gamma}$ and
the second basis does likewise for  $\hat{J}_{cl}$, $\hat{J}_{\alpha}$ 
and $\hat{J}_{\gamma}$. Hence, we can write the eigenstate
$|J_{cl},m_{\alpha},m_{\gamma}\rangle$ in terms of the eigenstates
$|j,m_{\alpha},m_{\gamma}\rangle$:
\begin{equation}
|J_{cl},m_{\alpha},m_{\gamma}\rangle
=\sum_{j}C_{j,J_{cl}}|j,m_{\alpha},m_{\gamma}\rangle,
\label{sup}
\end{equation}
where $C_{j,J_{cl}}$ are complex coefficients that depend only on
$j$ and $J_{cl}$.
To put it another way,
any  $|J_{cl},m_{\alpha},m_{\gamma}\rangle$ is a superposition
of states $|j,m_{\alpha},m_{\gamma}\rangle$ with the same $m_{\alpha}$
and $m_{\gamma}$ but different $j$.

Since we are only interested in the eigenvalues $\{J_{cl}\}$,
let us restrict ourselves to the normalized eigenfunctions 
\begin{equation}
\Psi _{J_{cl},0,0}(\alpha,\beta,\gamma)
\equiv \langle \alpha,\beta,\gamma|J_{cl},0,0\rangle,
\end{equation}
where  $m_{\alpha}$ and $m_{\gamma}$ have been
set to zero for convenience. This enables us to write 
\begin{equation}
\hat{J}_{cl}^{2}\Psi_{J_{cl},0,0}
=\hat{J}_{\beta}^2\Psi_{J_{cl},0,0}
=-\hbar^2{\partial^2\over \partial\beta^2}\Psi_{J_{cl},0,0}, 
\end{equation}
where  Eq.(\ref{41}) has also been incorporated.
Moreover, since $\Psi_{J_{cl},0,0}$
is an eigenfunction of $\hat{J}_{cl}$, it follows that 
\begin{equation}
-{\partial^2\over \partial\beta^2}\Psi_{J_{cl},0,0}
=J_{cl}^{2}\Psi_{J_{cl},0,0}.
\end{equation}

Inspecting the above equation, we are able to deduce 
the following:
\begin{eqnarray}
\Psi_{J_{cl},0,0} & \sim & \cos(J_{cl}\beta)\;\;\;{\rm with}\;\;\;
J_{cl}=0,1,2,...\nonumber\\
\Psi_{J_{cl},0,0} & \sim & \sin(J_{cl}\beta)\;\;\;{\rm with}\;\;\;
J_{cl}=1/2, 3/2, 5/2,...
\label{aba}
\end{eqnarray}
where  the identification $\beta +\pi=\pi -\beta$
\cite{EUL} has been employed.
It is, essentially, this 
identification of the Euler angel $\beta$ that constrains
$J_{cl}$ in the above manner. However, this is not yet the full
story because, as stressed above, any state  
$|J_{cl}\rangle$ can be written as a superposition
of states $|j\rangle$ (with the other, redundant labels
having been suppressed). It just so happens that  $\Psi_{j,0,0}$
is a symmetric function of $\beta$~\cite{EUL}  and, therefore, 
$\Psi_{J_{cl},0,0}$ must  be as well. On this basis, we can discard
the lower line in Eq.(\ref{aba}); thus restricting
$J_{cl}$ to strictly integer values. Moreover, we will find further support
for this restriction below. (Also, one might intuitively argue
that such an intrinsically classical form of angular momentum
should  be constrained in precisely this way.)

Combining the above outcome with Eqs.(\ref{39},\ref{40}),
we finally have an explicit expression for the
area spectrum of a rotating black hole:
\begin{equation}
A_{n,J_{cl}} =8\pi\hbar\left(n+  J_{cl}+{1\over 2}\right),\quad\quad\quad
n,{J_{cl}}=0,1,2,....
\label{46}
\end{equation}
This formulation for the area spectrum is the main result of the paper. 
Significantly, we have found the spectrum to be evenly spaced,
with the importance of this feature having been stressed
in the introductory section.

That the quantum number $J_{cl}$ should be restricted
to taking on  integer  values can
also be seen, independently of the above considerations,
  by way of the following discussion. 
Before elaborating on the logistics, let us 
point out that the same argument will provide some
further motivation for the periodicity conjecture
of Eq.(\ref{24}).

Firstly, it is useful to consider, in  the coordinate representation
with ${\hat J}_{cl}=-i\hbar\partial/\partial {\cal P}_{cl}$,
the  wavefunctions for the  angular-momentum
eigenstates. That is:
\begin{equation}
\Psi_{J_{cl}}({\cal P}_{cl})\sim \exp\left[i {J}_{cl}{\cal P}_{cl}
\right], 
\label{500}
\end{equation}
where  $J_{cl}$ is, as before, the eigenvalue
of $\hat{J}_{cl}/\hbar$; however, for the moment, we are assuming
no knowledge with regard to this spectrum.
In view of this formulation, we can make the following identification:
\begin{equation}  
 {J}_{cl}{\cal P}_{cl}\sim  {J}_{cl}{\cal P}_{cl} +2\pi p,
\label{501}
\end{equation}
where $p$ is an arbitrary integer. 

Next, let us recall Eq.(\ref{32}). Also employing
the explicit form of $B$ (\ref{37}) and the first law of
black hole mechanics (\ref{1}), we can elegantly re-express
this relation as follows:
\begin{equation} 
{\cal P}_{cl}=\chi+\theta,
\label{502}
\end{equation}
where we have defined $\chi\equiv \Pi_{cl}+\Omega \Pi_{M}$ and 
$\theta\equiv\kappa \Pi_{M}$. When $\chi$ is held
 constant, then  
 Eqs.(\ref{501},\ref{502}) tell us 
that $\theta$ should be constrained according to:\footnote{One
might be concerned that we are treating  $\chi$ and $\theta$
as independent variables,
whereas both depend on the conjugate $\Pi_{M}$. However,
$\chi$ also depends on a variable, $\Pi_{cl}$, which is clearly independent
of $\Pi_{M}$. Hence, we can, without loss of generality,
restrict ourselves to the case in which 
 $\chi$ is held constant.}
\begin{equation}
J_{cl}\theta \sim J_{cl}\theta+ 2\pi p;
\label{503}
\end{equation}
that is, $J_{cl}\theta$ must be an angle.
However, $\theta$ is, itself, an angle by hypothesis 
({\it cf}, Eq.(\ref{24}));
and so Eq.(\ref{503}) really says that $J_{cl}$ must
be strictly an integer,  thus reconfirming
our prior finding. Alternatively,
we could have used the spectrum of $\hat{J}_{cl}$ and Eq.(\ref{503})
to argue that $\theta$  should be an angle, thus supporting
the periodicity constraint (\ref{24}) via independent
means.

Although our work here is essentially done, one
important question remains: how does the ``classical''
spin eigenvalue, $J_{cl}$, relate to the more
conventional spin eigenvalue, $j$? As will be shown below,
$J_{cl}\approx j$ for $j\gg 1$.

To establish our claim, we  begin  by using Eq.(\ref{49}) to
evaluate $\langle j, m_{\alpha}, m_{\gamma}|
\hat{J}_{cl}^{2}-\hat{J}^{2}|j, m_{\alpha}, m_{\gamma}\rangle$.
It can be seen from the inverted form of Eq.(\ref{sup}) that this expectation 
value is independent of both $m_{\alpha}$ and $m_{\gamma}$.
Hence, we  denote it by $<\hat{J}_{cl}^2-\hat{J}^2>_{j}$ and, 
without loss of generality,  make the convenient choice of
$m_{\alpha}= m_{\gamma}=j$, for which the wavefunction is 
known~\cite{EUL}:
\begin{equation}
\Psi_{j,j,j}= {(i)^j\over 2\pi}\sqrt{{2j+1\over 2}}
\cos^{2j}\left({\beta\over 2}\right)\exp\left[-ij(\alpha
+\gamma)\right]
\label{50}
\end{equation}
and is normalized as follows:
\begin{equation}
\int^{2\pi}_0 d\alpha\int^{\pi}_0 d\beta
\int^{2\pi}_0 d\gamma \Psi^{*}\sin\beta\Psi=1.
\label{51}
\end{equation}

Directly applying the above formalism and Eq.(\ref{49}), we  obtain
the following:
\begin{eqnarray}
<\hat{J}_{cl}^2-\hat{J}^2>_{j} &=&
\hbar^2\int^{2\pi}_0 d\alpha\int^{\pi}_0 d\beta
\int^{2\pi}_0 d\gamma \Psi_{j,j,j}^{*}\cos\beta{\partial\over
\partial\beta}\Psi_{j,j,j}
\nonumber\\
&=& -{\hbar^2\over 2}\left(j-{1\over 2}\right).
\label{52}
\end{eqnarray}
Moreover, since $<\hat{J}^2>_j=\hbar^2 j(j+1)$, it follows that
\begin{eqnarray}
<\hat{J}_{cl}^2>_j&=&\hbar^2\left(j^2+{j\over 2}+{1\over 4}\right)
\nonumber\\
&\sim& j^2+{\cal O}[j].
\label{53}
\end{eqnarray}
This means that, for the physically interesting case of
$J_{cl}>>1$, we have  $J_{cl}\sim j$ and the area spectrum (\ref{46})
simplifies to
\begin{equation}
A_{n, j} \sim 8\pi\left(n+ j\right).
\label{54}
\end{equation}

A related point of interest is the mass spectrum
of the rotating black hole. In principle, this spectrum is
obtainable by way of Eqs.(\ref{3},\ref{46}).
Here, we focus on the  regime of large $J_{cl}\sim J$  and take
note of the following cases:
\begin{equation}
<{\hat M}>\sim \sqrt{{n\over 2}} \quad\quad if\quad n>>J>>1, 
\label{55}
\end{equation}
\begin{equation}
<{\hat M}>\sim \sqrt{J} \quad\quad if \quad J>>n>>1,
\label{56}
\end{equation}
\begin{equation}
<{\hat M}>\sim{1\over 2}\sqrt{5 n} \quad\quad if \quad J\sim n>>1.
\label{57}
\end{equation}

Finally, let us consider the ``inverse'' of the 
calculation in Eq.(\ref{53}); that is, 
$$
\langle J_{cl}, m_{\alpha}, m_{\gamma}|\hat{J}^2
|J_{cl}, m_{\alpha}, m_{\gamma}\rangle\equiv
<\hat{J}^2>_{J_{cl}}.
$$
It follows from prior considerations that this
expectation value should indeed be independent of
$m_{\alpha}$ and $m_{\gamma}$. Hence, we can make this evaluation
for the particularly simple case of $m_{\alpha}=m_{\gamma}=0$.
Incorporating $\Psi_{J_{cl},0,0}\sim \cos(J_{cl} \beta)$
({\it cf}, Eq.(\ref{aba})), into the same general
framework as depicted in Eq.(\ref{52}), we 
find that $<\hat{J}_{cl}^2-\hat{J}^2>_{J_{cl}}$ is
identically vanishing. 
In view of this outcome, it directly follows that
\begin{equation}
<\hat{J}^2>_{J_{cl}}=<\hat{J}_{cl}^2>_{J_{cl}}=J_{cl}^2.
\label{59}
\end{equation}
Therefore, when the system  is expressed in
terms of the unorthodox (but completely legitimate)
set of eigenstates  $|J_{cl},m_{\alpha},m_{\gamma}\rangle$,  
the operators  $\hat{J}$ and
$\hat{J}_{cl}$  are effectively indistinguishable. 

\section{Conclusion}

In summary, we have studied the area spectrum of a
rotating (Kerr) black hole in four dimensions of spacetime.
Extending a treatment by Barvinsky {\it et al}~\cite{BDG},
we have  demonstrated that the area spectrum
is evenly spaced, as it depends exclusively
on a pair of integer-valued
quantum numbers. To quantize the spin sector, 
 we have applied a novel approach
that utilizes the Euler components of the classical
angular momentum. We have shown that the operator form
of this classical angular momentum -
which is closely
related to but nevertheless distinct from the ``conventional''
quantum spin operator - has a spectrum of eigenvalues
that is restricted to integer  values. 
We have shown that,  when the angular momentum
is large (as expected to be the case for a
physically realistic black hole), this spectrum
is in asymptotic agreement with the ``usual''
quantum spin number, $j$.  We have also demonstrated
that quantum fluctuations prevent extremal black holes from
appearing in the physical spectrum.
Notably, an analogous censoring mechanism has already
been found for the case of charged (but static) black holes~\cite{BDG}.

Let us again point out that our approach
incorporates a pair of conjectural inputs. Firstly,
we have assumed that a rotating black hole
can be described in terms of  a reduced phase space,
consisting of  a ``handful''
of physical observables and their respective conjugates.
In view of prior works on static black holes
\cite{KUN,KUC,LMG}, this appears to be a reasonable 
assumption, but one that  should still 
be formally addressed. Secondly, we
have followed~\cite{BDG} in assuming that the 
canonical conjugate to the mass
is periodic, with the period fixed in accordance with 
purely Euclidean considerations. This seems difficult
to establish on a rigorous level, but appears  intuitively
correct when one considers that the Euclidean sector
plays a fundamental role in the very notion of
black hole thermodynamics~\cite{GH}.  We have also
provided  support for this periodicity condition by
way of an independent argument.

It would be interesting to compare our outcomes with that
of a prior, related  work by Makela {\it et al}~\cite{MRLP}.
However, because of a discrepancy with regard to precisely what
quantity is being quantized - $A-A_{ext}$ for us versus
$A+A_{-}$ for them\footnote{Here, 
 $A_{-}$ represents the  area of the {\it inner}
black hole horizon, whereas $A_{ext}$ describes
the area of either horizon at extremality.
 See Section 1 for further details.} -
a direct comparison would be highly non-trivial. 
Nonetheless, one might expect that  the {\it qualitative} features
of the spectrum
persevere for the case of large angular momentum,
and this does indeed seem to be the case.
\par
Finally, let us comment on the possibility of future directions.
One might naively expect that extending the analysis to include
charge  would be trivial; however, this is not quite
correct, as we will fully elaborate on in an upcoming paper \cite{NEW}.
 Meanwhile, a change in the spacetime dimensionality would 
involve technical
difficulties (one would require the higher-dimensional analogues
of the Euler components), but should be straightforward in principle.
Another interesting problem would be to relate our findings
to those of other studies, such as the  surface quantization  approach
of  Khriplovich~\cite{KHR} or the hyperspin formalism advocated
by one of the authors~\cite{GOUR}. We defer such intrigue until
a future time.

\section{Acknowledgments}

The authors would like to thank V.P. Frolov for
helpful conversations. GG is also grateful for the Killam Trust
for its financial support.

\end{document}